\documentclass[osajnl,twocolumn,showpacs,superscriptaddress,10pt]{revtex4-1}
\usepackage{amsmath,amssymb,graphicx}
\begin{document}

\title{Frequency comb generation beyond the Lugiato-Lefever equation:
  multi-stability and super cavity solitons}

\author{Tobias Hansson}
\email{tobhan@chalmers.se}
\affiliation{Department of Applied Physics, Chalmers University of
  Technology, SE-41296 G\"oteborg, Sweden}
\affiliation{Dipartimento di Ingegneria dell'Informazione, Universit\`a di Brescia, via Branze 38, 25123 Brescia, Italy}
\author{Stefan Wabnitz}
\affiliation{Dipartimento di Ingegneria dell'Informazione, Universit\`a di Brescia, via Branze 38, 25123 Brescia, Italy}

\begin{abstract}
The generation of optical frequency combs in microresonators is considered
without resorting to the mean-field approximation. New dynamical
regimes are found to appear for high intracavity power that cannot be modeled using
the Lugiato-Lefever equation. Using the Ikeda map we show the existence of multi-valued
stationary states and analyse their stability. Period doubled patterns are
considered and a novel type of super cavity soliton associated with the multi-stable states is predicted.
\end{abstract}


\maketitle

\section{Introduction}
\label{sec:intro}

The generation of optical frequency combs using microresonator devices
is currently a hot topic that is attracting a significant amount of
research. A mode-locked optical frequency comb can combine
an ultra broadband spectrum with coherent and finely resolved comb
lines at an equidistant frequency spacing. Microresonator based
optical frequency combs have a plethora of potential applications in,
e.g., the areas of precision metrology, spectroscopy,
optical clocks, wavelength-division multiplexing sources, and arbitrary
waveform generation \cite{kippenberg_microresonator-based_2011}.
By enclosing a nonlinear medium within a resonator structure, a wealth
of dynamical behaviours appear. Microresonators with a Kerr
nonlinearity that are pumped by a continuous wave (CW) laser are known
to support bistable CW solutions, as well as both periodic temporal (Turing)
patterns and localised cavity soliton solutions. These solutions may be either stable or
unstable, depending on the operating regime determined by the
power and detuning of the pump laser. Additionally, there exist operating regimes where
the intracavity field is chaotic.

Previous work on the generation of optical frequency combs in
microresonator devices has primarily focused on the regime where the intracavity
power is relatively low. In this case the mean-field approximation
is valid, and the dynamics can be accurately modeled using either the
Lugiato-Lefever equation (LLE) \cite{lugiato_spatial_1987} or using the theory of coupled mode
equations \cite{coen_modeling_2013,chembo_modal_2010}.
For these approaches to be applicable, it is assumed that the overall cavity detuning, including
both linear and nonlinear contributions, remains small during
propagation. Particularly, it is assumed that the characteristic nonlinear length scale is much
longer than the path length of the cavity. In the present work, we aim to go beyond the
LLE and the mean-field approximation, and
consider the resonator dynamics when the nonlinear phase-shift is
relatively large (of the order of unity).

As a motivation for why it might be prudent to consider microresonator
dynamics beyond the LLE, we note that mean-field models have been
unable to account for the complete range of dynamical
behaviours that have been experimentally observed in resonator devices. Recent experimental
work at NIST has e.g.~shown examples of frequency combs that are not
well described by the LLE \cite{delhaye_phase_2015}. Mean-field models are also
unable to account for the period doubling behaviour, which has been
observed to occur in dissipative fiber-ring resonators \cite{nakatsuka_observation_1983,temporal_valle_1991,coen_experimental_1998}.
This period doubling behaviour is distinct from the periodic breathing of cavity
soliton type solutions of the LLE, which occurs on a different
time-scale that is not associated with the roundtrip time. Period doubling in fiber-ring
cavities has been experimentally observed in the periodic switching of
the amplitude of output pulses when a fiber-ring resonator is pumped by a
synchronised pulse train with large input amplitude \cite{temporal_valle_1991}.
The period doubling sequence is also well known to be an essential component in the route to
chaos \cite{nakatsuka_observation_1983,Ankiewicz_chaos_1987}.

While the lengths of microresonators are generally about four or so orders of
magnitude shorter than fiber-ring resonators, they can display
nonlinearities that are several orders of magnitude greater than the
$\chi^{(3)}$ nonlinearity of silica fiber. This is especially true for microring resonators
made of semiconductor materials. The use of highly nonlinear microring
resonators with fairly long path lengths, in combination with intense
input pump powers, could consequently enable the observation of
microresonator dynamics similar to what has previously been observed in fiber-ring resonators.

In the next section \ref{sec:model} we introduce the Ikeda map, which is the basic
model system used to describe frequency comb generation beyond the
LLE. We use the Ikeda map to analyse the CW solution in
section \ref{sec:stat} and show that additional stationary states with
higher intracavity power appear when compared
to the LLE. The stability of these stationary states is investigated in section \ref{sec:stability}, showing the presence of
new parametric instabilities other than the conventional modulational
instability for high pump powers. In Section \ref{sec:P2pattern} we use the parametric instability to show
examples of period doubling behaviour. Thereafter, in section \ref{sec:CS}, we
demonstrate the intriguing possibility of forming super energetic cavity soliton associated with the excited
states. Finally, in the last section \ref{sec:conclusions}, we summarise the article and
present some overall conclusions.

\section{The Ikeda map}
\label{sec:model}

In this section we introduce the Ikeda map \cite{ikeda_multiple-valued_1979},
which we use to model the dynamics of optical frequency combs without resorting to the
mean-field approximation. The map is constructed by combining
an ordinary nonlinear Schr\"odinger equation, for describing the
evolution of the envelope of the electric field within a waveguide,
together with boundary conditions that relate the fields between
successive roundtrips and the input pump field
\cite{haelterman_dissipative_1992,coen_modeling_2013}, viz.
\begin{align}
  & E^{m+1}(t,0) = \sqrt{\theta}E_{in} +
  \sqrt{1-\theta}e^{i\phi_0}E^m(t,L), \label{eq:Ikeda1}\\
  & \frac{\partial E^m(t,z)}{\partial z} = -\frac{\alpha_i}{2}E^m(t,z)
  - i\frac{\beta_2}{2}\frac{\partial^2 E^m(t,z)}{\partial t^2} \nonumber\\
  & + i\gamma|E^m(t,z)|^2E^m(t,z). \label{eq:Ikeda2}
\end{align}
Here the independent variables are the evolution variable $z$, which
is the longitudinal coordinate measured along the waveguide and $t$ which is the
(ordinary) time. The first equation is the boundary condition that
determines the intracavity field $E^{m+1}(t,z=0)$ at the input of
roundtrip $m+1$ in terms of the field from the end of the previous
roundtrip $E^m(t,z=L)$ and the pump field $E_{in}$. The path length of
the resonator is assumed to be equal to $L$. Additionally, $\theta$ is the
coupling transmission coefficient and $\phi_0 = 2\pi l-\delta_0$ is
the linear phase-shift, with $\delta_0$ the frequency detuning from the cavity
resonance closest to the pump frequency, assumed to
correspond to the longitudinal mode number $l = 0$. Eq.(\ref{eq:Ikeda2}) is written in the
reference frame moving at the group velocity, with
second-order group velocity dispersion coefficient $\beta_2$ and nonlinear
coefficient $\gamma = n_2\omega_0/(c A_{eff})$ (with $n_2$ being the
material Kerr coefficient, $\omega_0$ the angular pump frequency, $c$ the
speed of light in vacuum and $A_{eff}$ the effective mode area), as well a linear
absorption coefficient $\alpha_i$. In this work we will for simplicity
neglect higher-order dispersion and any frequency dependence of the
nonlinear coefficient. However, since Eq.(\ref{eq:Ikeda2}) is simply
the evolution equation of a straight waveguide it can trivially be
extended to include any higher-order effects.

In the usual treatment of microresonator frequency combs,
cf.~\cite{matsko_mode-locked_2011,coen_modeling_2013},
one proceeds by averaging the map
Eqs.({\ref{eq:Ikeda1}-\ref{eq:Ikeda2}}) over one roundtrip in
order to obtain the LLE
\begin{align}
  & t_R\frac{\partial E(t,\tau)}{\partial\tau} +
  i\frac{\beta_2L}{2}\frac{\partial^2 E(t,\tau)}{\partial t^2}
  -i\gamma L|E(t,\tau)|^2E(t,\tau) = \nonumber\\
  & -(\alpha + i\delta_0)E(t,\tau) + \sqrt{\theta}E_{in},
  \label{eq:LLE}
\end{align}
where the evolution variable $z$ has been replaced by the slow time
$\tau$. The coefficient $t_R$ is the roundtrip time and $\alpha = (\alpha_i
L+\theta)/2$ is the total cavity loss.

Since the LLE can be derived from the Ikeda map, the latter is
consequently more general. The Ikeda map is easy to simulate
using conventional numerical methods for the nonlinear Schr\"odinger
equation such as the split-step Fourier method \cite{hansson_on_2014}.
However, numerical simulations of the Ikeda map must use a step length
that is shorter than the cavity length, which makes it more
computationally demanding than the LLE.

The derivation of the mean-field approximation is valid provided that the total cavity
detuning is much less than unity, i.e.~$\gamma L |E|^2 \ll
1$ and $\delta_0 \ll 1$, see Ref.~\cite{haelterman_dissipative_1992}.
The latter condition signifies that the pump detuning
from resonance is small, while the former states that the nonlinear
phase-shift is also a small quantity, which can be equivalently interpreted as a requirement
for the nonlinear length scale $L_{nl} = 1/(\gamma |E|^2)$ to be much longer than the cavity
circumference. The average dispersion over each roundtrip should additionally remain
small. Note, however, that this does not necessarily imply that the field must be
changing slowly inside the cavity, cf.~\cite{conforti_modulational_2014}.
Below we will investigate some of the new dynamical regimes that
appear as the nonlinear length scale becomes comparable to the cavity
length, i.e. $L_{nl} \approx L$.

\section{Multi-valued stationary solutions}
\label{sec:stat}

In this section we consider continuous wave solutions of
the Ikeda map Eq.(\ref{eq:Ikeda1}-\ref{eq:Ikeda2}). Particularly, we
will consider CW solutions where the intracavity field can
decay along the resonator circumference due to intrinsic absorption
($\alpha_i > 0$), but where the same amplitude is reproduced after each
roundtrip with the coherent addition of the pump field. Fields that
are periodically restored can be considered stationary in the
averaged sense of the mean-field theory.

\begin{figure}[ht]
\centerline{\includegraphics[width=.65\columnwidth]{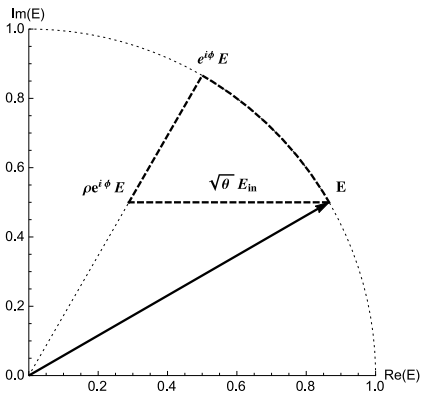}}
\caption{Geometric illustration of a fixed point of Eq.(\ref{eq:Fixedp}). The
  original vector is reproduced after applying a rotation, a scaling
  and adding the pump vector. Either one or three stationary states
  are possibile when the rotation (phase-shift) is assumed small, but
  additional stationary states can also be obtained if the rotation is
  by more than one lap.}
\label{fig:Fixedp}
\end{figure}

The CW solution of the nonlinear Schr\"odinger equation (\ref{eq:Ikeda2}) that accounts
for the field that has propagated around the resonator is given by $E^m(t,L) =
E^m(0)\exp(-\alpha_i L/2 + i\gamma L_{eff}|E^m(0)|^2)$, with $L_{eff}
= (1-e^{-\alpha_i L})/\alpha_i$ being the effective nonlinear length
due to intrinsic absorption ($L_{eff} \to L$ as $\alpha_i \to
0$). Assuming that $E^{m+1}(0) = E^m(0) = E$ we find that the intracavity
field should satisfy the following fixed point equation
\begin{equation}
  E = \rho e^{i\phi}E + \sqrt{\theta}E_{in}
  \label{eq:Fixedp}
\end{equation}
which has a simple geometrical interpretation, see
Fig. \ref{fig:Fixedp} and Ref.~\cite{Ankiewicz_chaos_1987}. Meanwhile,
the intracavity power must satisfy the following condition
\begin{equation}
  \frac{1}{(1-\rho)^2 + 4\rho\sin^2(\phi/2)} = \frac{|E|^2}{\theta|E_{in}|^2}
  \label{eq:CW}
\end{equation}
\begin{equation}
  \phi = \delta_0 - \gamma L_{eff} |E|^2
  \nonumber
\end{equation}
where the roundtrip loss coefficient $\rho =
\sqrt{1-\theta}e^{-\alpha_i L/2} \approx 1 - \alpha$.
The right-hand-side of Eq.(\ref{eq:CW}) has a linear dependence on the
intracavity power $|E|^2$ while the left hand side can be recognised as an Airy
function describing resonances whenever $\phi = 2\pi n$ with $n$ an integer.
\begin{figure}[ht]
\centerline{\includegraphics[width=.8\columnwidth]{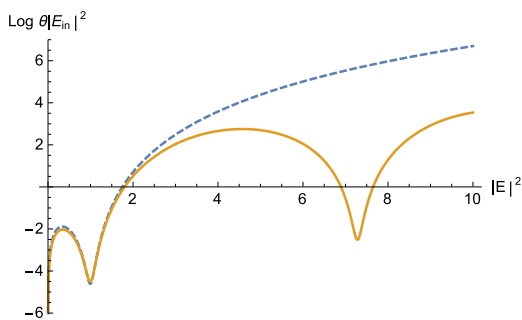}}
\caption{Comparison of pump dependence on
  the intracavity power for multi-valued stationary continuous
  wave solutions of the LLE (blue dashed) and the Ikeda map
  (orange solid). Parameters $\alpha = \theta = 0.1$ and $\gamma = L = P_{in}
  = \delta_0 = 1$.}
\label{fig:Levels2}
\end{figure}

The corresponding bistable steady-state condition for the LLE (\ref{eq:LLE}) is
given by
\begin{equation}
  \frac{1}{\alpha^2 + (\delta_0-\gamma L |E|^2)^2} = \frac{|E|^2}{\theta|E_{in}|^2}
%  \theta|E_{in}|^2 = |E|^2\left[(\delta_0-\gamma L |E|^2)^2+\alpha^2\right].
  \label{eq:LLEstat}
\end{equation}
A comparison of Eqs.(\ref{eq:CW}-\ref{eq:LLEstat}) is shown in
Fig.~\ref{fig:Levels2}, where it is seen that the LLE (blue dashed curve) is an excellent approximation for
small intracavity powers but that it quickly becomes bad as the
power increases. Indeed, from Eq.(\ref{eq:CW}) we see that the intracavity
power will have a linear asymptotic dependence on $|E|^2$, while Eq.(\ref{eq:LLEstat}) for the LLE predicts that the
intracavity power should depend asymptotically on the cubic root of the pump power.
The intensity inside the resonator will consequently increase much faster
than the LLE predicts. From Fig.~\ref{fig:Levels2} we also see that
higher-order multi-stable (or excited) states do not necessarily
imply that the pump power needs to be large. In fact one finds that
excited solutions may even coexist with the low-power bistable response.

The successive intracavity powers for which the pump power has a
minimum in Fig.~\ref{fig:Levels2}, i.e., the least pump power needed to
reach a state on a given level, can be found explicitly from Eq.(\ref{eq:CW}) as $|E|^2 =
(\delta_0 + 2\pi n)/(\gamma L)$, where $n$ is a positive
integer, while the corresponding pump powers are given by $\theta
|E_{in}|^2 = (\delta_0+2\pi n)(1-\rho)^2/(\gamma L)$. The pump power
required to reach the first excited level is thus a fraction $1 +
2\pi/\delta_0$ larger than that required in order to reach the upper
level of the bistability curve. Therefore, the observation of excited
states is facilitated if the detuning is fairly large. In this article we refer to
states with $n \geq 1$ as excited states, since the LLE is only
able to model the first minimum occurring for $n = 0$.

We now consider the physical parameters necessary to obtain a
nonlinear phase-shift of order unity in microresonator devices. The phase condition
$\phi_{NL} = \gamma L P_0 \sim 1$ depends on the intracavity power $P_0 = |E|^2$
rather than the pump power. However, from Eq.(\ref{eq:CW}) one can
estimate that the power will grow linearly with the resonator finesse,
i.e.~$P_0 \sim P_{in} \mathcal{F}/\pi$. This approximation corresponds
to a lower bound for a critically coupled resonator.
The phase-shift condition can alternatively be written in terms of the
resonator Q-factor and the effective mode area as
$\phi_{NL} \approx 2\frac{n_2}{n_0}\frac{P_{in}Q}{A_{eff}} \sim 1$.
The nonlinear phase-shift is related to the Kerr frequency mode shift as
$\Delta\nu_{NL} = \phi_{NL}\Delta\nu_{FSR}/(2\pi)$, with a phase-shift of
$2\pi$ thus able to shift the field to a different resonance.
Note that the modes can also experience a thermally induced
mode shift, however, this is not taken into account in the current treatment.

Microresonators featuring large mode shifts have in fact already been
experimentally demonstrated in the literature. For example, in
Ref.~\cite{delhaye_octave_2011} a fused silica microtoroid, with a
diameter of $80~\mu m$ and $Q \approx 2.7\times 10^8$, was used to demonstrate octave spanning
comb generation. This resonator had a Kerr frequency shift that was
estimated to be as much as $100~GHz$ at $1~W$ of input power, while the FSR of the resonator was $850~GHz$.
The corresponding nonlinear phase-shift would therefore be $\phi_{NL}
\approx 0.74$, which suggests that the mean-field approach may be
inaccurate. The comb bandwidth of the same resonator was also
theoretically analysed in Ref. \cite{coen_universal_2013}, with
experimental measurements showing deviations of one order of magnitude from predictions
based on the LLE. These large deviations were however attributed to
be mainly due to higher-order dispersion.

\section{Linear stability analysis}
\label{sec:stability}

Optical frequency comb generation in microresonators pumped by a CW
laser is initiated by the growth of frequency sidebands above a
certain threshold power. These sidebands are symmetrically spaced
and occur due to modulational instability (MI) of the pump mode.
Because of the cavity boundary conditions, MI in a resonator cavity
can display qualitatively different features, compared to the conventional
MI in a straight waveguide. This is because the detuning between the
resonance frequency of the cavity and the pump frequency, introduces
an additional degree of freedom for the phase-matching of different
four-wave-mixing processes. As a consequence, the steady-state CW solution
can become modulationally unstable also in the normal dispersion
regime \cite{haelterman_additive-modulation-instability_1992}.

An analysis of the linear stability of the steady-state solution in
the mean-field limit has been performed by many authors, see
e.g.~Refs. \cite{matsko_optical_2005,chembo_modal_2010,hansson_dynamics_2013}.
However, the LLE does not model the full dynamics of the instabilities that can occur
in a resonator cavity described by the Ikeda map. Particularly, one
finds that additional unstable frequency bands (resonance tongues)
appear at high power. These bands can be associated both with
instabilities having the fundamental period and with period doubling
instabilities \cite{coen_modulational_1997}.

In order to investigate the linear stability of steady-state solutions
described by Eq.(\ref{eq:CW}), we assume an ansatz of the form
$E^m(t,z) = [E_0 + u^m(t,z) + i v^m(t,z)]\exp[-\alpha_i z/2 + i\gamma (1-e^{-\alpha_i
z}) |E_0|^2/\alpha_i + i Arg~E_0]$, where the perturbations $u^m(t,z)$ and $v^m(t,z)$
are real functions. Linearising the nonlinear Schr\"odinger Eq.(\ref{eq:Ikeda2}) we find
that the Fourier transform of the perturbations $w^m(z) \equiv
[\tilde{u}^m(\omega,z),~\tilde{v}^m(\omega,z)]^T$ should satisfy the following system of linear
equations $dw^m/dz = M w^m$, with the perturbation matrix
\begin{equation}
  M(z) = \left[\begin{array}{cc}
    0 & -(\beta_2/2)\omega^2 \\
    (\beta_2/2)\omega^2 + 2\gamma |E_0|^2 e^{-\alpha_i z} & 0
    \end{array}\right].
  \label{eq:Mmatrix}
\end{equation}
The analytical solution of this system is quite involved in the general case and
the stability is most easily investigated numerically.
However, in the absence of intrinsic absorption, i.e.~$\alpha_i \to
0$, the matrix $M$ becomes independent of $z$ and the solution
simplifies considerably, see
Refs.~\cite{coen_modulational_1997,zezyulin_modulational_2011}.
In this case one finds that the eigenvalues
of $M$ give the conventional MI gain of a straight waveguide, viz.
\begin{equation}
  \mu = \omega\sqrt{-\beta_2\gamma |E_0|^2 - (\beta_2^2/4)\omega^2},
  \label{eq:gamma}
\end{equation}
while the solution of the linear system is given by
\begin{equation}
  w^m(z) =
  \left[\begin{array}{cc}
      -\beta_2\omega^2e^{\mu z}  & -\beta_2\omega^2e^{-\mu z} \\
      2\mu e^{\mu z} & -2\mu e^{-\mu z} \end{array}\right]
  \left[\begin{array}{c} a_m\\ b_m \end{array}\right] \equiv W(z) c^m
\end{equation}

In the next step we consider the boundary condition
Eq.(\ref{eq:Ikeda1}) that the perturbed solution must satisfy and
apply the Fourier transform. This leads to $w^{m+1}(0) =
\sqrt{1-\theta}R(\phi)w^m(L)$, where $\phi = \phi_0 + \gamma L
|E_0|^2$ and $R(\phi)$ is the rotation matrix
\begin{equation}
  R(\phi) = \left[\begin{array}{cc}
    \cos\phi & -\sin\phi\\
    \sin\phi & \cos\phi
  \end{array}\right].
\end{equation}
We then obtain a system of difference equations
for the coefficients $c^m = [a_m,~b_m]^T$, viz. $c^{m+1} =
\sqrt{1-\theta}[W^{-1}(0)R(\phi)W(L)] c^m \equiv Q c^m$.
The stability of solutions to this linear system of difference equations with
constant coefficients depends, similarly to Eq.(\ref{eq:Mmatrix}), on
the magnitude of the eigenvalues. The Ikeda map has an instability
whenever the modulus of an eigenvalue of $Q$ is larger than unity.

For the case when the intrinsic absorption is neglected these
eigenvalues are explicitly given by
\begin{equation}
  q_\pm = \sqrt{1-\theta}(p \pm \sqrt{p^2-1})
\end{equation}
where $p = \cos(\phi)\cosh(\mu L)-\left[\gamma|E_0|^2+(\beta_2/2)\omega^2\right]\sin(\phi)\sinh(\mu
L)/\mu$, with $\mu$ given by Eq.(\ref{eq:gamma}) and
the phase-shift $\phi = \phi_0 + \gamma L |E_0|^2$, see Ref.~\cite{coen_modulational_1997}.

For the general case the stability of the Ikeda map can be determined
using a numerical Floquet analysis \cite{conforti_modulational_2014,bender_advanced_1999}.
This is accomplished by investigating the stability of eigenvalues of the fundamental matrix
$[w_1^{m+1}(0),~w_2^{m+1}(0)]$. This matrix is obtained by numerically integrating
the perturbed system $dw^m/dz = M(z) w^m$ over one roundtrip, starting
from two linearly independent initial conditions, e.g.~$w_{1,2}^m(0) = [1,~0]^T,
[0,~1]^T$, and applying the boundary condition Eq.(\ref{eq:Ikeda1}).

\begin{figure}[ht]
\centerline{\includegraphics[width=\columnwidth]{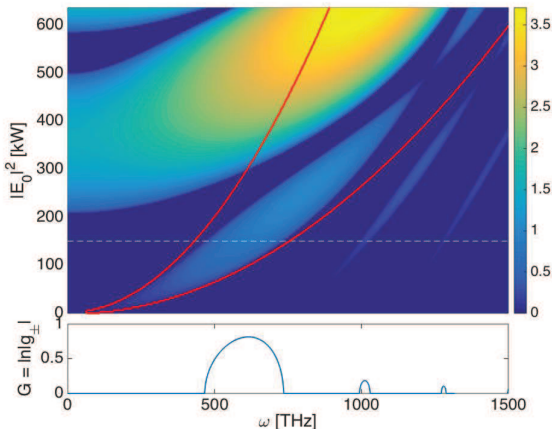}}
\caption{Parametric instability tongues of the Ikeda map for anomalous
dispersion. The red contour shows the predicted range of
modulational instability for the LLE. Below a cross section corresponding to
the dashed line is shown with alternating CW-MI/P2-MI gain
bands. Parameters: $\gamma = 25~W^{-1}km^{-1}$,  $L = 2\pi\times 40~\mu m$, $\mathcal{F} =
10^6$, $\beta_2 = -40~ps^2km^{-1}$ and $\delta_0 = 0$.}
\label{fig:anomalousMI}
\end{figure}

\begin{figure}[ht]
\centerline{\includegraphics[width=\columnwidth]{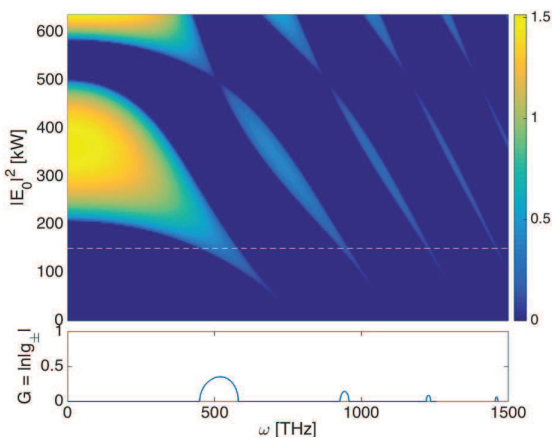}}
\caption{Parametric instability tongues of the Ikeda map for the same
  parameters as in Fig. \ref{fig:anomalousMI} but with the sign of the
  dispersion reversed to give normal dispersion $\beta_2 =
  40~ps^2km^{-1}$. The cross section shows alternating P2-MI/CW-MI
  gain bands. Note that the LLE does not predict any instability for these parameters.}
\label{fig:normalMI}
\end{figure}

In Figs.~\ref{fig:anomalousMI}-\ref{fig:normalMI} we show some
some examples of modulational instability gain for both anomalous and
normal dispersion. The gain has been calculated numerically using the Floquet method
for parameters corresponding to the silica resonator of
Ref.~\cite{delhaye_octave_2011} mentioned at the end of section
\ref{sec:stat}. Note that even though the magnitude of the
intracavity power is very large, the maximum plotted value may
actually correspond to pump powers as low as $2~W$, owing to the very high finesse
$\mathcal{F} \approx 10^6$ of the resonator.
Fig.~\ref{fig:anomalousMI} shows that the modulational instability
analysis for the LLE is valid for low intracavity power, but that new unstable sidebands
appear as the power increases. The high power parametric instability
tongues may correspond to instabilities occurring for either resonant or anti-resonant
conditions. The former includes the ordinary CW-MI of the LLE and higher
order sidebands, and has its origin in the $q_+$ eigenvalue, while the
latter comes from the $q_-$ eigenvalue and gives rise to a
period doubling instability referred to as P2-MI, see
Refs.~\cite{mclaughlin_new_1985,coen_modulational_1997}.
The P2-MI can be used to convert a continuous wave solution to a
period doubled pattern, i.e.~an alternation between two different
modulated temporal patterns having a period of one roundtrip
each. Since an optical spectrum analyser performs an
averaging over many roundtrips, it may be difficult to experimentally
distinguish a stable period doubled frequency combs
from an ordinary frequency comb.

While the LLE gives a good approximation for the modulational
instability gain occurring at low powers, its accuracy becomes increasingly bad as the
power increases. Indeed, the LLE is not able to model either P2-MI instabilities nor
higher order CW-MI sidebands. Fig.~\ref{fig:normalMI} shows that the
period doubling instability is in fact usually the first instability to occur for normal dispersion
cavities, except for the case when the detuning has been used to
compensate for the phase-mismatch. Moreover, we see that high
intensity continuous wave solution may actually be stable in certain
narrow regions.

The excited continuous wave steady-state solutions generally
correspond to the parameter regions which are stable to plane wave perturbations. However, the stability
analysis shows that the new states are usually modulationally
unstable (CW-MI and/or P2-MI) for anomalous dispersion, similar to the upper steady-state level
in the simple bistable case of the LLE. But since the modulational
instability only applies to structures that are sufficiently broad in
time, it suggests the intriguing possibility of forming cavity solitons associated with these
excited multi-stable states. Indeed, this turns out to be possible
under appropriate circumstances and we shall see an example of such
excited cavity solitons in section \ref{sec:CS}.

\section{Period doubled patterns}
\label{sec:P2pattern}

In this section we give some examples of period doubled patterns.
A period doubled CW solution should satisfy a system of
equations similar to that of a steady-state fixed point solution,
viz.
\begin{equation}
  A = \rho e^{i\phi_B}B + \sqrt{\theta}E_{in}, \qquad B = \rho e^{i\phi_A}A + \sqrt{\theta}E_{in}.
  \label{eq:Fixedp2}
\end{equation}
Here we have defined the half-period fields as $A = E^m$ and $B =
E^{m+1}$. The presence of a period doubled fixed point implies that
the field satisfies a functional relation of the form $x = F(F(x))$. If the fields in each
half-period are equal, i.e.~$A = B = E$ (or $x = F(x)$), there is no period doubling
and we recover Eq.(\ref{eq:Fixedp}). From Eq.(\ref{eq:Fixedp2}) one
finds that the two fields are related to each other through the
following resonance condition, see Ref.~\cite{haelterman_period-doubling_1992},
\begin{equation}
  \frac{A}{1 + \rho e^{i\phi_B}} = \frac{B}{1 + \rho e^{i\phi_A}},
  \label{eq:P2cond}
\end{equation}
where $\phi_{A,B}$ denotes the phase-shift with the nonlinear term
evaluated for $A$ or $B$, respectively.

Period doubled states can occur for relatively low intracavity powers
$|A|^2$ and $|B|^2$ if the cavity detuning is close to anti-resonant conditions, i.e.~$\phi
\approx \pi$. The above condition (\ref{eq:P2cond}) also shows that period doubling
cannot occur in a linear medium since the denominators cancel each
other in the absence of a nonlinear phase-shift.

Two coupled conditions for the intracavity power can be derived, viz.
\begin{equation}
  \theta|E_{in}|^2 = |A|^2\frac{(1-\rho^2)^2 +
    4\rho^2\sin^2\left[(\phi_A+\phi_B)/2\right]}{(1+\rho)^2 -
    4\rho\sin^2(\phi_B/2)}
\end{equation}
for $A$ and an analogous condition for $B$ which is obtained by interchanging
$A$ and $B$ in the above equation. After solving these equations, the fields can be recovered using the relation $A =
\sqrt{\theta}E_{in}(1+\rho e^{i\phi_B})/(1-\rho^2e^{i(\phi_A+\phi_B)})$.

\begin{figure}[ht]
\centerline{\includegraphics[width=\columnwidth]{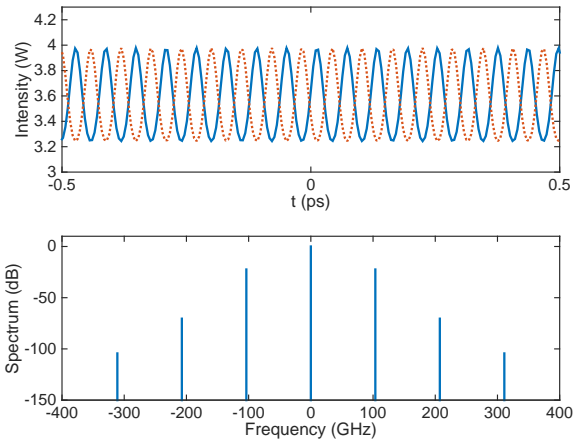}}
\caption{Period doubled pattern obtained through P2-MI
  instability for normal dispersion resonator. Two identical but out of phase patterns alternate between
each roundtrip (blue solid line denotes even numbered roundtrips, red dotted
odd). Numerical simulation of the Ikeda map using parameters $\alpha =
\theta = 0.01$, $\gamma = 100~W/m$, $L = 0.1~mm$, $P_{in} = 4.5~W$,
$FSR = 100~GHz$, $\beta_2 = 500~ps^2/km$ and $\delta_0 = 0.25$.}
\label{fig:P2-pattern}
\end{figure}

Period doubled CW solutions are usually only obtained for large input
pump powers in microresonator systems where the finesse is high. However, other
period doubled patterns can be found at rather modest power levels as a result of period doubling
instabilities. An example is shown in Fig.~\ref{fig:P2-pattern} for a
microresonator with normal cavity dispersion. The period doubled pattern
consists of two identical but out of phase patterns that alternate
between each roundtrip. The pattern is found to be stable for the
considered operating parameters, but it becomes unstable due to the growth
of additional sidebands if the pump power is increased further. Note
that the spectrum is constant for the current example since
only the spectral phase changes between each roundtrip.

\section{Localised solutions, cavity solitons}
\label{sec:CS}

An important question that we have not considered so far is whether
or not the Ikeda map has any additional types of localised solutions
different from those of the LLE. That the answer to this question should be in the
positive can be appreciated from the following line of argument.

It is well known that cavity solitons exist for almost the same range of
parameters as the bistable CW solution in microresonators described
by the LLE. The reason for this is that a bright cavity soliton consists of a solitary pulse
sitting on top of a constant background. This background is
a stationary and modulationally stable solution of Eq.(\ref{eq:CW}). Similarly,
since cavity solitons are concave at their peak, there should also exist
another stationary CW solution that corresponds to two points on the
soliton were the second derivative is zero. However, it should be noted that this
is only approximately true, due to the phase difference between the soliton
and the background. The two points are inflexion points, i.e.~points on the soliton profile were the
slope of the derivative goes from increasing to decreasing or vice
versa. The latter stationary CW level is however not required to be stable against
MI as long as the soliton is not too broad. Consequently, it can be
understood that one should be in a regime were at least two
CW solutions coexist for the same pump parameters in order to find a
stationary cavity soliton solution. Similarly, if more than two stationary CW states are present
simultaneously, then we can expect that each new level may also be
associated with a different cavity soliton state.

\begin{figure}[ht]
\centerline{\includegraphics[width=.8\columnwidth]{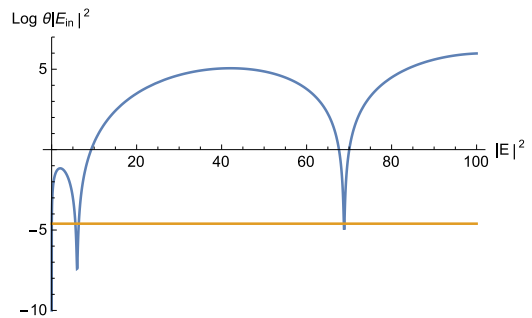}}
\caption{Configuration for stationary CW solutions allowing for multi-level cavity solitons. Logarithm of pump power plotted vs.~intracavity
power. The constant line corresponds to the external pump power ($1~W$) used in the examples.}
\label{fig:CS-level}
\end{figure}

\begin{figure}[ht]
\centerline{\includegraphics[width=\columnwidth]{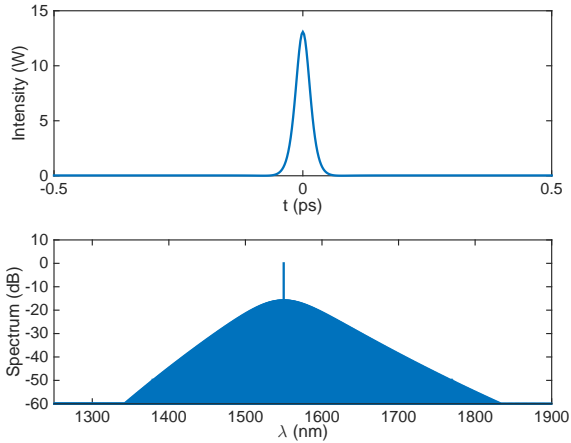}}
\caption{LLE type cavity soliton with background level $|E|^2 \approx
  0.029~W$ and support level $|E|^2 \approx 6.365~W$. Simulation of Ikeda
  map Eqs.(\ref{eq:Ikeda1}-\ref{eq:Ikeda2}), with parameters $\alpha = \theta = 0.01$, $\gamma = 100~W/m$, $L =
0.1~mm$, $P_{in} = 1~W$, $FSR = 100~GHz$, $\beta_2 = -500~ps^2/km$ and
$\delta_0 = 0.6$.}
\label{fig:CS1}
\end{figure}

\begin{figure}[ht]
\centerline{\includegraphics[width=\columnwidth]{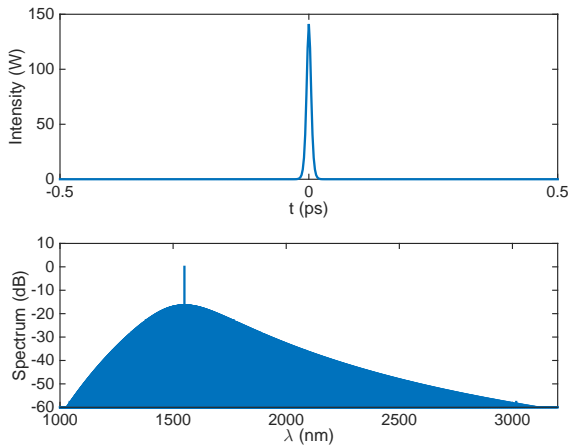}}
\caption{Stationary super cavity soliton with background level $|E|^2 \approx
  0.029~W$ and support level $|E|^2 \approx 68.764~W$, using the exact same parameters
as in Fig. \ref{fig:CS1} except for different initial conditions.}
\label{fig:CS2}
\end{figure}

To find a cavity soliton associated with an excited CW state we
consider operating parameters allowing for three simultaneous
steady-state CW levels with positive slope, see Fig.~\ref{fig:CS-level}.
For the numerical simulations we consider a microresonator with parameters $\alpha = \theta = 0.01$, $\gamma = 100~W/m$, $L =
0.1~mm$, $FSR = 100~GHz$, $\beta_2 = -500~ps^2/km$, $P_{in}  = 1~W$ and $\delta_0 = 0.6$. These parameters are somewhat
idealised since we neglect higher order dispersion, but similar
parameters could be obtained for semiconductor
microrings featuring large nonlinear Kerr coefficients such as
e.g.~AlGaAs. In Figs.~\ref{fig:CS1}-\ref{fig:CS2} we demonstrate a case were
two different stable cavity solitons, one conventional LLE soliton and
one super cavity soliton associated with
an excited state, are obtained and may coexist simultaneously inside
the cavity.
The two stationary solitons in Figs.~\ref{fig:CS1}-\ref{fig:CS2} are
found in the same microring using identical pump parameters.
The second cavity soliton that is associated with the
excited CW level around $|E|^2 \approx 68.8~W$ in
Fig.~\ref{fig:CS-level} is seen to be much more energetic than its
LLE counterpart, that is associated with the CW power
level $|E|^2 \approx 6.4~W$, corresponding to the upper bistable CW
solution of the LLE. The soliton shown in Fig.~\ref{fig:CS2} has a peak power that
is one order of magnitude greater than that of the LLE
soliton. At the same time it is also narrower, leading to a
mode-locked frequency comb spectrum that is significantly broader than
that of the LLE soliton. This is particularly interesting for
applications to broadband frequency comb generation, since such super cavity
solitons could easily span a bandwidth of more than an octave, and
possibly even multiple octaves with appropriate dispersion
engineering. Other potential applications include multi-state optical
buffer memories, cf.~\cite{leo_temporal_2010}.

The intensity of each soliton sans background can be approximated as
$|E(t)|^2 = P_0\textrm{sech}^2(t/T_0)$, with peak intensity $P_0 = 2(\delta_0+2\pi n)/(\gamma
L)$ and temporal duration $T_0^2 = \beta_2 L/(2(\delta_0+2\pi n))$ for
$n = 0,1$. Note that the two solitons were excited numerically by starting with initial conditions
close to the final soliton profile. This corresponds experimentally
to the injection of a soliton pulse into the cavity. At present it is not clear how
super cavity solitons could best be excited in an experimental
setting. This and the relation of super cavity solitons to other types
of bistable solitons, cf.~\cite{kaplan_bistable_1985}, will be the focus of future work.

\section{Conclusions}
\label{sec:conclusions}

In this article we have presented an analysis of frequency comb generation that goes beyond
the mean-field approximation used in the Lugiato-Lefever equation and
coupled mode theory. The analysis has been carried out in the context of
microresonators but is valid also for dispersive fiber-ring resonators.
The LLE is valid only in the limit of small phase-shifts
were the nonlinear length scale is much longer than the length of the
cavity. Using the more general Ikeda map as a model, we have shown that new
multi-valued stationary states appear at high intracavity powers for
the CW solution. These states may be present simultaneously as the
low power bistable response and do not necessarily require a large
pump intensity. The modulational stability of the stationary CW states
has been analysed, while taking into account the
boundary conditions imposed by the cavity. The stability analysis
shows the presence of several parametric instability bands, including period
doubling instabilities, at high intracavity power both for normal and
anomalous dispersion. An example of how such period doubling
instabilities can lead to a stable pattern that alternates between each
roundtrip has been demonstrated. Finally, we have predicted a novel type of
super cavity soliton solution associated with an excited CW
steady-state. These solitons are narrower and more energetic than
those of the LLE, allowing for efficient generation of mode-locked
ultra wideband frequency combs.

\section*{Acknowledgements}

This research was funded by Fondazione Cariplo (grant no.~
2011-0395), the Italian Ministry of University and Research
(grant no.~2012BFNWZ2) and by the Swedish Research Council (grant
no.~2013-7508). S.W. is also with Istituto Nazionale di Ottica (INO)
of the Consiglio Nazionale delle Ricerche (CNR).

\end{document}